\documentclass[longauth]{aa}

\usepackage{graphicx}
\usepackage[varg]{txfonts}
\begin{document}

   \title{ALMA sub-mm maser and dust distribution of VY Canis Majoris}

 \author{A.~M.~S.~Richards\inst{1}\fnmsep\thanks{\email{\tt amsr@jb.man.ac.uk}}
          \and          C.~M.~V.~Impellizzeri\inst{2,4}
          \and          E.~M.~Humphreys\inst{3}
          \and         C.~Vlahakis\inst{4}
          \and          W.~Vlemmings\inst{5}
          \and          A.~Baudry\inst{6,7}
          \and          E.~De~Beck\inst{5}
          \and          L.~Decin\inst{8}
          \and          S.~Etoka\inst{9}
          \and          M.~D.~Gray\inst{1}
          \and          G.~M.~Harper\inst{10}
          \and          T.~R.~Hunter\inst{2}
          \and          P.~Kervella\inst{11,12,13}
          \and          F.~Kerschbaum\inst{14}
          \and          I.~McDonald\inst{1}
          \and          G.~Melnick\inst{15}
          \and          S.~Muller\inst{5}
          \and          D.~Neufeld\inst{16}
          \and          E.~O'Gorman\inst{5}
          \and          S.~Yu.~Parfenov\inst{17}
          \and          A.~B.~Peck\inst{2}
          \and          H.~Shinnaga\inst{18}
          \and          A.~M.~Sobolev\inst{17}
          \and          L.~Testi\inst{3}
          \and          L.~Uscanga\inst{19}
          \and          A.~Wootten\inst{2}
          \and          J.~A.~Yates\inst{20}
          \and          A.~Zijlstra\inst{1} 
}
\institute{JBCA, School of Physics and Astronomy, Univ. of Manchester, UK\label{1}
\and NRAO, 520 Edgemont Road, Charlottesville, VA22903, USA\label{inst2}
\and ESO Karl-Schwarzschild-Str. 2, 85748 Garching, Germany\label{inst3}
\and Joint ALMA Observatory/European Southern Observatory, Alonso de Cordova 3107, Vitacura, Santiago, Chile\label{inst4}
\and Department of Earth and Space Sciences, Chalmers University of Technology, Onsala Space Observatory, SE 439 92 Onsala, Sweden\label{inst5}
\and    Univ. Bordeaux, LAB, UMR 5804, F-33270 Floirac, France\label{inst6}
\and   CNRS, LAB, UMR 5804, F-33270 Floirac, France\label{inst7}
\and Instituut voor Sterrenkunde, Katholieke Universiteit Leuven, Celestijnenlaan 200D, 3001 Leuven, Belgium\label{inst8}
\and  Hamburger Sternwarte, Univ. of Hamburg, Gojenbergsweg 112, D-21029 Hamburg,  Hamburg, Germany\label{inst9}
\and  School of Physics, Trinity College Dublin, Ireland\label{inst10}
\and LESIA, Observatoire de Paris, CNRS, UPMC, Universit\'e Paris-Diderot, PSL, 5 place Jules Janssen, 92195 Meudon, France\label{inst11}
\and UMI-FCA, CNRS/INSU, France (UMI 3386)\label{inst12}
\and Dept. de Astronom\'{\i}a, Universidad de Chile, Santiago, Chile\label{inst13}
\and Dept. of Astrophysics, Univ. of Vienna, T\"{u}rkenschanzstra{\ss}e 17, 1180, Vienna, Austria\label{inst14 }
\and  Harvard-Smithsonian Center for Astrophysics, 60 Garden Street, MS 66, Cambridge, MA 02138, USA \label{inst15}
\and Dept. of Physics \& Astronomy, Johns Hopkins Univ. 3400 North Charles Street, Baltimore, MD 21218, USA\label{inst16}
\and Ural Federal University, Ekaterinburg, Russia\label{inst17 }
\and  NAOJ, 2-21-1 Osawa, Mitaka, Tokyo, Japan zip 181-8588\label{inst18}
\and IAASARS, National Observatory of Athens, 15236 Athens, Greece\label{inst19}
\and Dept. of Physics and Astronomy, University College London, WC1E 6BT, UK\label{inst20}
  }
   \date{Received September 17, 2014; accepted October 16, 2014}

\abstract{}
{Cool, evolved stars have copious, enriched winds.  
 Observations have so far not fully constrained models for the shaping and acceleration of these winds.
We  need to understand the dynamics better, from the
pulsating stellar surface 
to $\sim$10 stellar radii,  where radiation pressure on dust is fully
  effective. Asymmetric nebulae around some red supergiants
imply the action of additional forces.}
{We retrieved ALMA Science Verification data providing images of
  sub-mm line and continuum emission from VY CMa. This enables us to
  locate water masers with milli-arcsec  accuracy and to resolve the dusty continuum.}
{The 658, 321, and 325 GHz masers lie in irregular, thick shells at
  increasing distances from the centre of expansion. For the first
  time this is confirmed as the stellar position, coinciding with a
  compact peak offset to the NW of the brightest continuum emission.
  The maser shells overlap but  avoid each
  other on scales  of up to 10 au. Their distribution is broadly
  consistent with excitation models but the conditions and kinematics
  are complicated by wind collisions, clumping, and asymmetries. }  {}

   \maketitle

\section{Introduction}
\label{Intro}

Massive stars have a profound impact on their surroundings via their
material and energy output.  Observations support the importance
  of radiation pressure on dust in driving the stellar wind, as
  reviewed by \citet{Habing96} and confirmed for red supergiants (RSG)
  by \citet{Mauron11}.  There are a variety of models for mass
  transport from the stellar surface to the dust formation zone at
  5--10 stellar radii ($R_{\star}$), based on combinations of
  convection \citep{Chiavassa11}, wind levitation by pulsation and dust
  formation (\citealt{Bowen88}; \citealt{Ireland06}), for example, including
  scattering as well as absorption \citep{Bladh13}. Acoustic and
  magnetic forces were analysed by \citet{Hartmann80}. However, observations do not yet confirm any of these as the dominant force.
  Oxygen-rich grain formation models have difficulty in explaining
  dust-driven winds \citep{Woitke06}, which can possibly be solved by the detection of large dust grains at a few $R_{\star}$ \citep{Norris12} and the
  complexity of the gas and dust distribution close to the photosphere
  \citep{Wittkowski07}. Moreover, most investigations have focused on
  low-mass stars, which have a different internal structure, and
  despite their high mass-loss rates, RSG have irregular, often
  shallow periods ({\small
    http://cdsarc.u-strasbg.fr/cgi-bin/afoevList?cma/vy}).  

We
investigated mass loss from VY CMa, one of the largest RSG, progenitor
mass $\sim$25 M$_{\odot}$, $R_{\star}$) 5.7 mas at 2$\mu$m
\citep{Wittkowski12}, at 1.2$\pm$0.1 kpc, \citealt{Choi08};
\citealt{Zhang12}).  It has had a high and variable mass-loss rate,
0.5--1$\times$10$^{-4}$ M$_{\odot}$ yr$^{-1}$ in its recent past
\citep{Decin06}, up to 3$\times$10$^{-3}$ M$_{\odot}$ yr$^{-1}$
\citep{Humphreys07}.
This provides the richest-known O-rich
circumstellar envelope (CSE) chemistry, as seen at sub-mm
wavelengths by \emph{Herschel} \citep{Alcolea13}, for instance, and imaged at
$\sim$1'' resolution using the SMA \citep{Kaminski13}.

   \begin{figure}
   \centering
   \includegraphics[width=7.9cm]{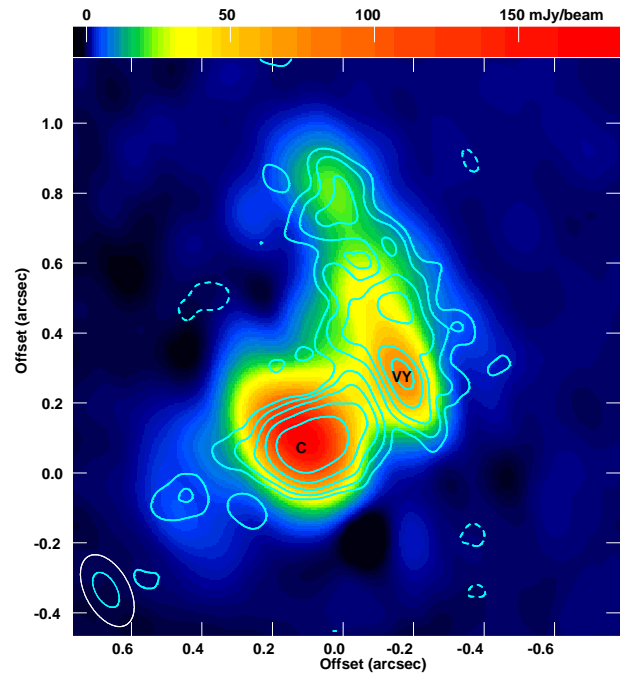}
   \caption{Continuum emission: 321 GHz colour scale,
     658 GHz contours at (--1,1,2,4,8,16)$\times$10 mJy
     beam$^{-1}$. Synthesized beams shown at lower left for 321 GHz (white), 658
     GHz (blue).  (0, 0) at R.A. 07 22 58.33454 Dec. --25 46
     03.3275 (J2000). {\bf C} marks the continuum peak. {\bf VY} is identified as the star, at the
     centre of the water maser expansion.  }
         \label{VYCMA_321_658_CONT.PS}
   \end{figure}

\onlfig{
   \begin{figure}
   \centering
   \includegraphics[width=9cm]{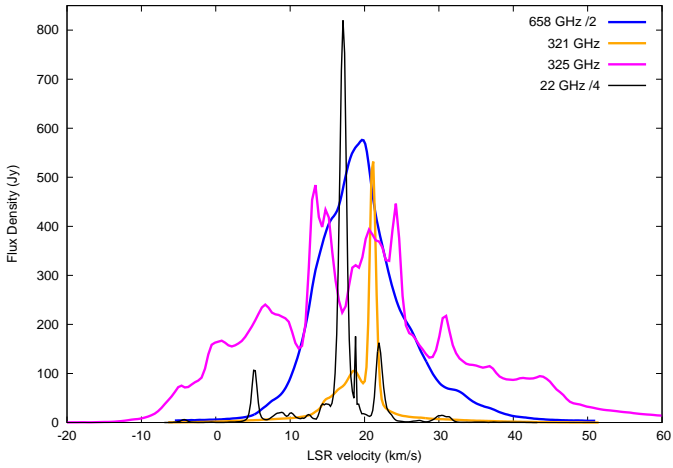}
   \caption{Integrated water maser spectra, derived from the
     interferometric image cubes, measured in square boxes of width
     $0\farcs75$, $1\farcs0$ and $0\farcs4$ at 321, 325, and 658 GHz,
     centred on {\bf VY}. The 658 GHz spectrum is scaled by 0.5 and
     the 22 GHz spectrum by 0.25.}
\label{Spectra_final.eps}
\end{figure}
}

   \begin{figure*}
   \centering
   \includegraphics[width=9cm]{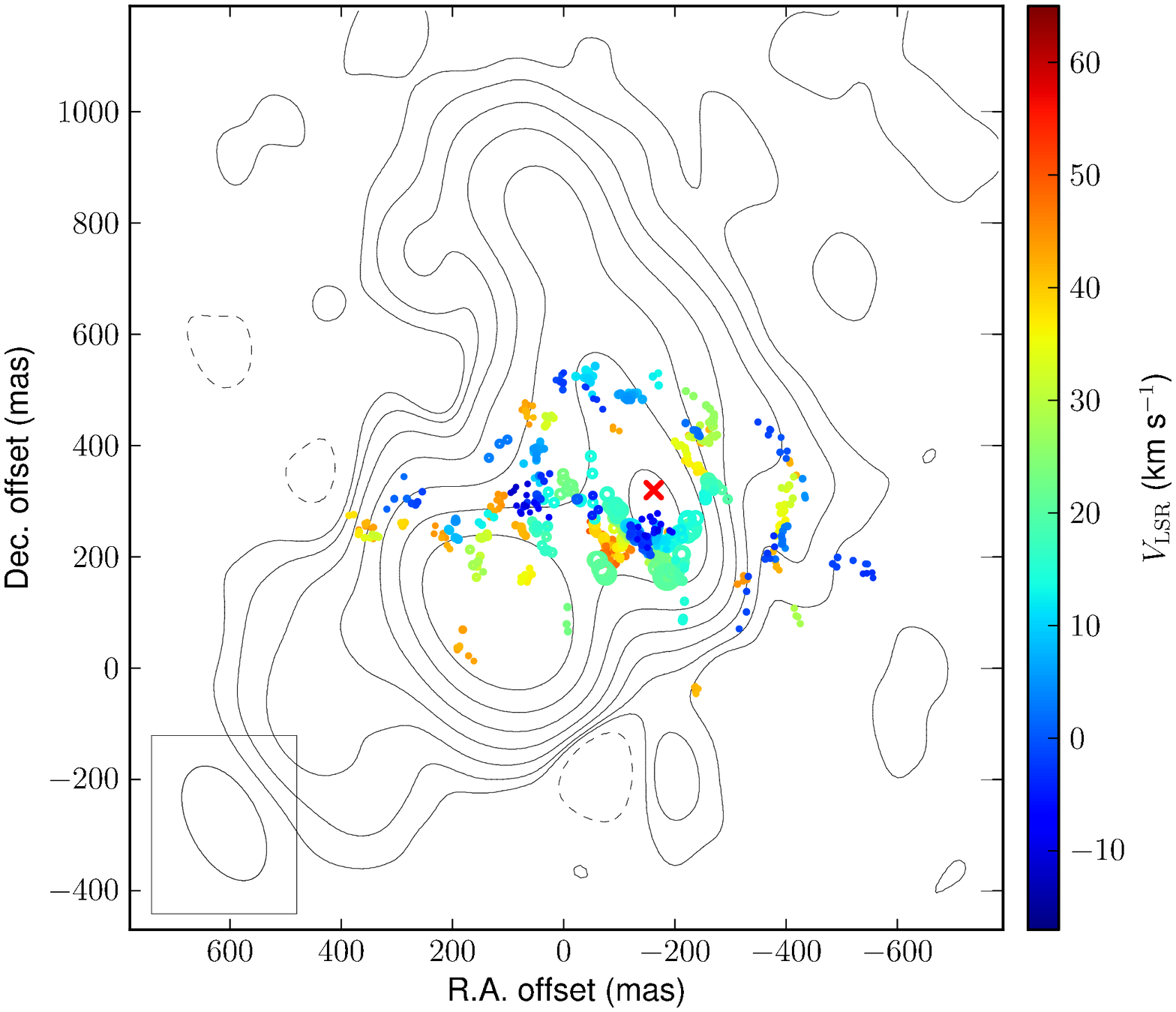}
   \includegraphics[width=9cm]{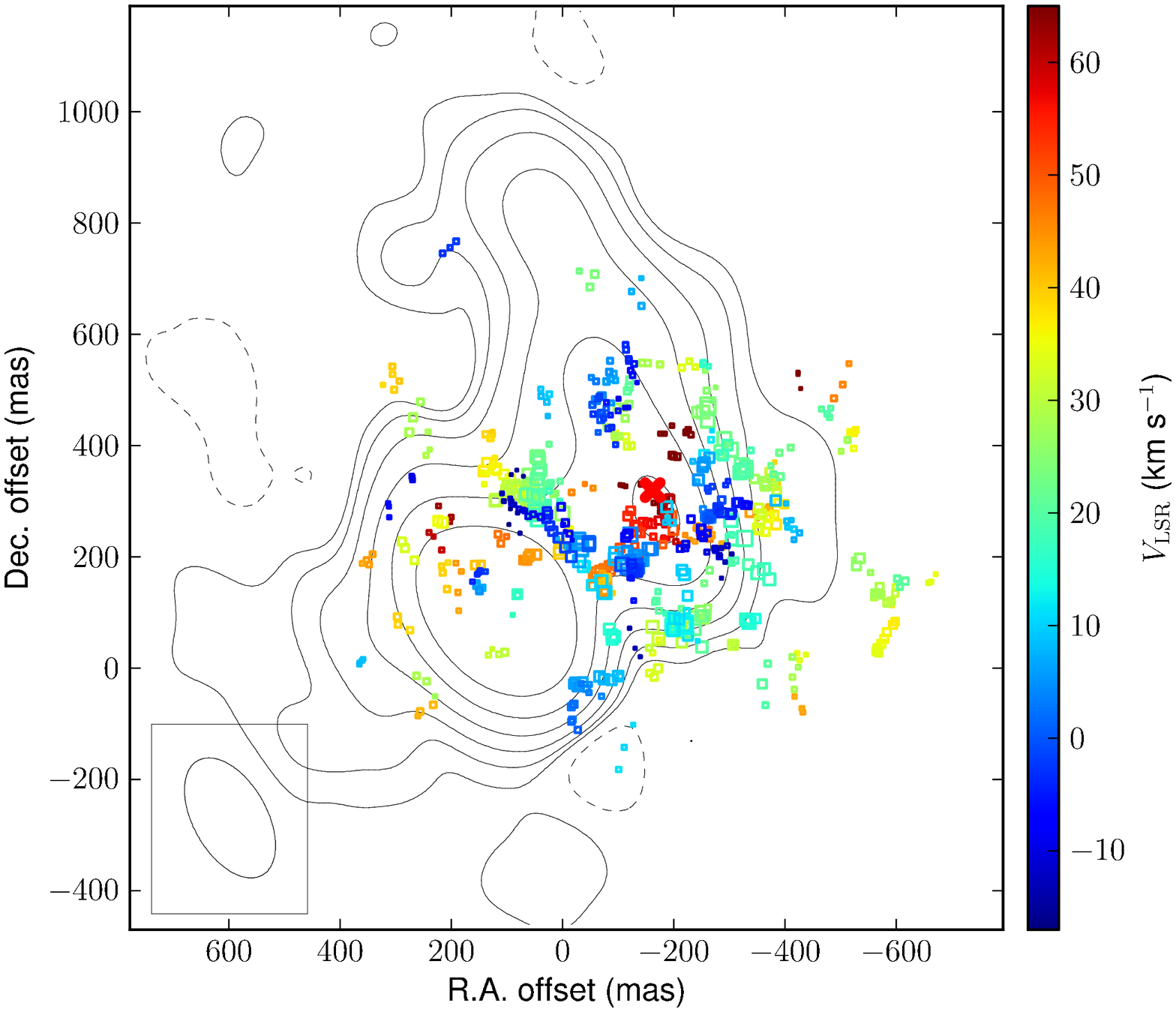}
\vspace*{-0.5cm}
\caption{321 GHz (left) and 325 GHz (right) maser positions over
  continuum contours, lowest levels at (--1, 1) and (--2) mJy
  beam$^{-1}$ at 321 and 325 GHz, respectively, and  at
  (2, 4, 8, 16, 32) mJy beam$^{-1}$ thereafter. Maser symbol size
  proportional to $\sqrt{\mathrm{flux; density}}$. The red cross marks {\bf VY}.}
         \label{321-325xyplots}
   \end{figure*}
   \begin{figure}
   \centering
   \includegraphics[width=9cm]{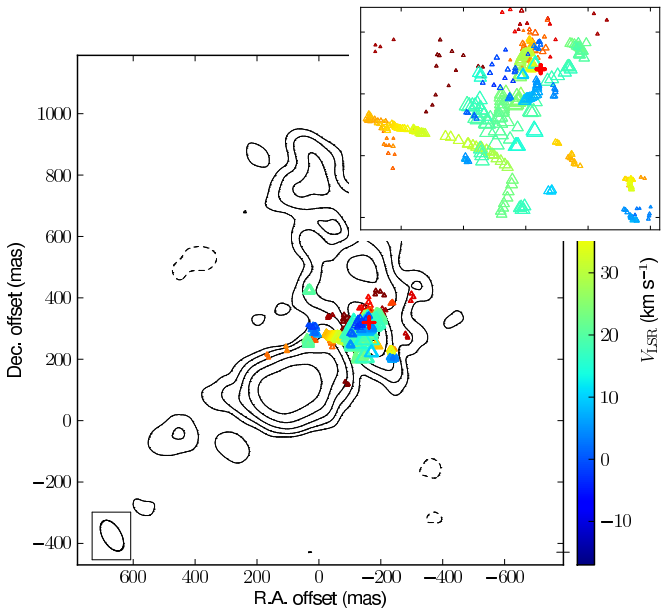}
\vspace*{-0.5cm}
\caption{658 GHz H$_{2}$O maser positions overlying continuum contours as in
  Fig.~\ref{VYCMA_321_658_CONT.PS}. Maser symbols and velocity scale as in
  Fig.~\ref{321-325xyplots}. The inset shows 
  the region with bottom left and top right  at (--20,
  190) and (--255, 370) mas.  }
         \label{658xyplots}
   \end{figure}

VY CMa has a highly asymmetric nebula that extends
over a few arcsec and is shaped like a
lopsided heart, irregular and clumpy on all scales
\citep{Humphreys07}.  VLA and SMA observations at 8.4--355 GHz show an
unresolved central ellipse, dominated by emission from dust,
(e.g. \citealt{Lipscy05}; \citealt{Kaminski13}).
Strong OH, SiO, and 22-GHz H$_{2}$O masers have been imaged by
many authors but, hitherto, there has been no astrometric confirmation
that the star lies at the centre of expansion.  The
22-GHz H$_{2}$O masers are located in a thick shell of radii 75--440
mas, with Doppler and proper motions dominated by accelerating outflow
\citep{Richards98V}. Their maximum expansion velocity is 35.5 km
s$^{-1}$ relative to the stellar velocity $V_{\star}$ of 22 km
s$^{-1}$ (all velocities are with respect to the local standard
of rest, LSR).

Models (\citealt{Gray12} and references therein; \citealt{Daniel13})
predict that 
 the 321.22564 GHz $J_{Ka,Kc}$ $10_{2,9}$--$9_{3,6}$, and possibly the
  325.15292 GHz  $5_{1,5}$--$4_{2,2}$ 
H$_{2}$O maser lines 
can emanate from conditions found at both sides of the dust
formation zone. Maser emission at 321 GHz needs hotter gas than at 22
GHz, whilst lower temperatures and number densities favour 325
GHz. This has been confirmed by imaging in Cepheus A \citep{Patel07},
but only the 22 GHz transition has ever been resolved in a CSE. The
 658.00655 GHz $v_2 = 1, 1_{1,0}$--$1_{0,1}$
GHz maser is expected to occur very close to the star under
conditions similar to SiO masers \citep{Hunter07}. More details of
these transitions are given in Table~\ref{tab:masers}. We present ALMA
observations that test these predictions and, for the first time,
resolve sub-mm masers, thermal lines, and continuum.

\section{Data acquisition and reduction}
\label{data}

We obtained public ALMA Science Verification data for VY CMa observed
on 2013 16--19 August using 16--20 12 m antennas on baselines from
0.014--2.7 km.  Three scheduling blocks (SB), covering each of the maser
lines, are referred to as the 321-GHz, 325-GHz and 658-GHz
SBs. 
The 
  velocity resolution after Hanning smoothing is 0.45 km s$^{-1}$ at
321 and 658 GHz, and 0.9 km s$^{-1}$ at 325 GHz.  More details of
observations and data reduction are given in Appendix B. 
The fully calibrated line-free channels were imaged
using a synthesized beam of (0\farcs22$\times$0\farcs13) at 321 and
325 GHz, and (0\farcs11$\times$0\farcs06) at 658 GHz.  After
subtracting the continuum, the masers were imaged using beam sizes of
(0\farcs18$\times$0\farcs09) and (0\farcs088$\times$0\farcs044) at
321/325 and 658 GHz.  We measured the positions of the
masers and continuum peaks by fitting Gaussian components using the
{\sc aips} task {\sc sad}, see Appendix B.

\section{VY CMa continuum and maser morphology}
\label{Continuum} Before self-calibration, the positions measured by
Gaussian fits to the 321 and 325 GHz peaks wandered by up to 35 mas,
but at 658 GHz there were offsets of up to 100 mas,
mainly due to differences between atmospheric conditions towards VY
CMa and the phase reference source (see Appendix B). The 321
GHz data,
with the best atmospheric transmission and sensitivity, were used for
astrometry and for continuum analysis around this frequency.
The continuum at all frequencies has a similar J shape, with a
bright extended peak {\bf C}, a secondary, compact peak {\bf VY} , and
several other peaks.  Shifts of
(--2, 0) and (--87, 27) mas were applied at 325 and 658 GHz,
respectively, to align {\bf VY} with its 321 GHz position, which
brought the other bright features into good positional agreement
(Fig.~\ref{VYCMA_321_658_CONT.PS}). The maximum detectable angular
extent in R.A. and Dec. is 1\farcs2$\times$1\farcs6 above the
3$\sigma_{\mathrm{rms}}$ contour at 321 GHz.  The 321 GHz position of
{\bf VY} is R.A. 07 22 58.3226 Dec. --25 46 03.043 (J2000), with
35 mas uncertainty  dominated by errors in transferring phase
  corrections from the reference source. {\bf VY} is 328$\pm$1
mas from {\bf C} at PA --33$^{\circ}$. Using a matching beam size
160$\times$64 mas$^{2}$, {\bf C} and {\bf VY} had peak flux densities
of 133.9 and 71.7 ($\sigma_{\mathrm{rms}}$ 0.9) mJy beam$^{-1}$ at 321
GHz and of 474 and 296 ($\sigma_{\mathrm{rms}}$ 4) mJy beam$^{-1}$ at
658 GHz. The continuum emission is analysed further by \citet{OGorman14}.

The total velocity extents of the H$_{2}$O lines are (--11.9 to 49.2),
(--16.8 to 75.1) and (--3.0 to 67.7) km s$^{-1}$ at 321, 325, and 658
GHz.
Figure~\ref{Spectra_final.eps} shows that the 325 GHz maser spectrum has
the largest  velocity extent and shows multiple strong peaks (similar to 22 GHz),
whilst the 321 and 658 GHz spectra are
dominated by single peaks, close to $V_{\star}$ at 321 GHz, but 2 km
s$^{-1}$ more blue-shifted at 658 GHz. Both these lines have a blue
shoulder. The 325  and 658 GHz masers have a long red tail.

We estimated the  brightness temperature $T_{\mathrm b}$ of each
  spatial component from its measured flux density $S$ (Appendix B)
and the beam size, that is,  $T_{\mathrm b}$ $\ge$756$S$ at 321 and 325
GHz, and $T_{\mathrm b}$ $\ge$3162$S$ at 658 GHz.  These give ranges
of (11--3.2$\times$$10^5$) K, (46--2.0$\times$$10^5$) K, and
(444--28$\times$$10^5$) K at 321, 325, and 658 GHz.  These values are
lower limits since maser component sizes are probably smaller
than the beam; $\le5$ mas components would increase
$T_{\mathrm b} \gg 1000$ K in all cases, for instance. However, some of the faintest
emission, especially at extreme velocities, may be genuinely extended,
thermal emission from H$_2$O or other species.

Figures~\ref{321-325xyplots} and~\ref{658xyplots} show that the masers
are clumped into spatially close groups at similar velocities.  Much
 of the 321 and 658 GHz emission is concentrated in  bright streamers
with clear velocity gradients, for example in Fig.~\ref{658xyplots} (insert),
where the long arc traces a continuum  `valley' between {\bf C} and
{\bf VY}. The 325 GHz masers are more evenly distributed. There is  more extended emission to the east of {\bf VY} in all
lines and brighter sub-mm masers to the south.  The 321 GHz masers
form a ring around {\bf VY}; the other lines have more blue- or
red-shifted emission along the line of sight to {\bf VY}
(Figs.~\ref{321-325xyplots}--\ref{VYCMa_22321325658_xyplot.eps}).  We estimated the centre of
expansion of each line as in \citet{Richards12}, by maximising the
separation between masers at close to $V_{\star}$, giving positions
very close to {\bf VY}.  Figure~\ref{VYCMa_321325658_ravel.eps} shows the
angular separation of each component from {\bf VY}, which appears to
be located in a maser-free sphere.
Figures~\ref{VYCMa_321325658_ravel.eps}
and~\ref{VYCMa_22321325658_xyplot.eps} show that the 658, 321, and
325 GHz maser inner rims are at successively larger distances from the
star, whilst the outer rims are at $\sim$ 200, 500, and 600 mas.

\onlfig{
   \begin{figure}
   \centering
   \includegraphics[width=9cm]{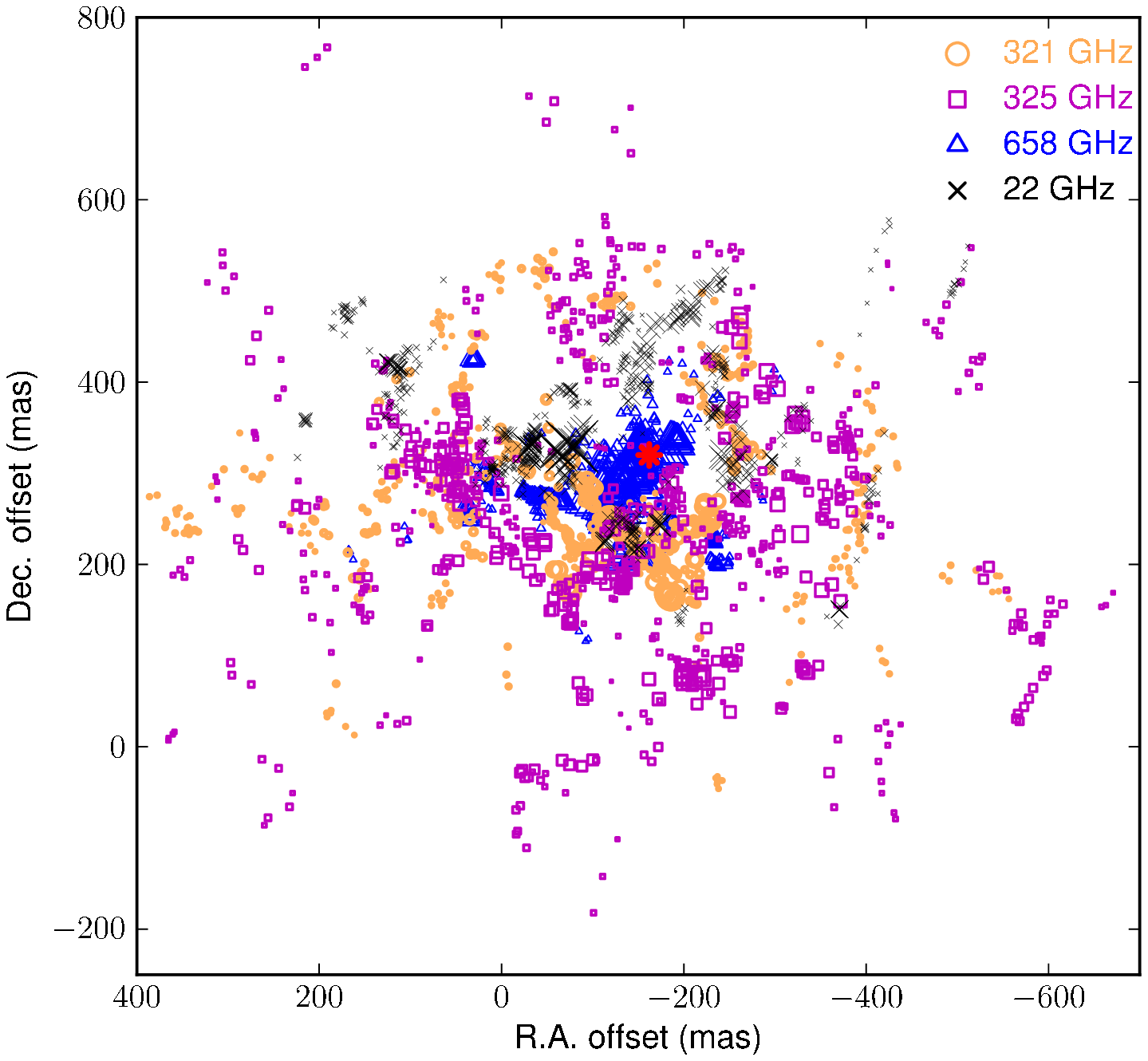}
\vspace*{-0.5cm}
\caption{Relative positions of all imaged maser components. }
         \label{VYCMa_22321325658_xyplot.eps}
   \end{figure}
}

   \begin{figure}
   \centering
   \includegraphics[width=9cm]{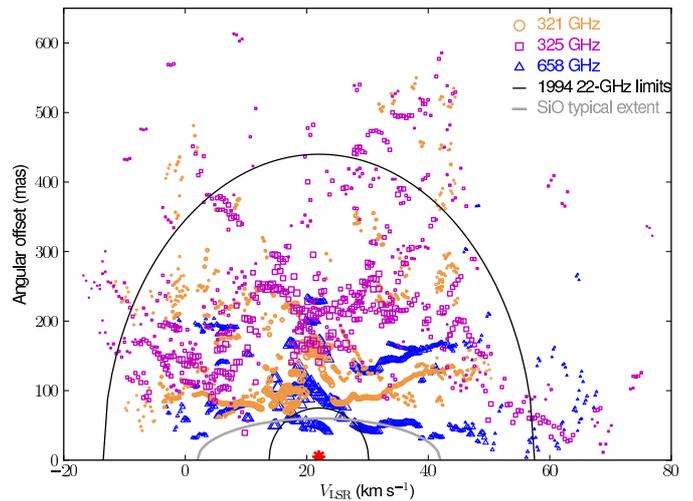}
\vspace*{-0.5cm}
\caption{Symbols mark sub-mm H$_2$O maser component angular separations from {\bf VY} as a
  function of $V_{\mathrm{LSR}}$. The black and grey lines mark the
  inner and outer limits of 22 GHz H$_2$O masers and the outer rim of
  J=1--0 and J=2--1 SiO masers.  }
         \label{VYCMa_321325658_ravel.eps}
   \end{figure}

\onlfig{
   \begin{figure}
   \centering
   \includegraphics[width=8cm]{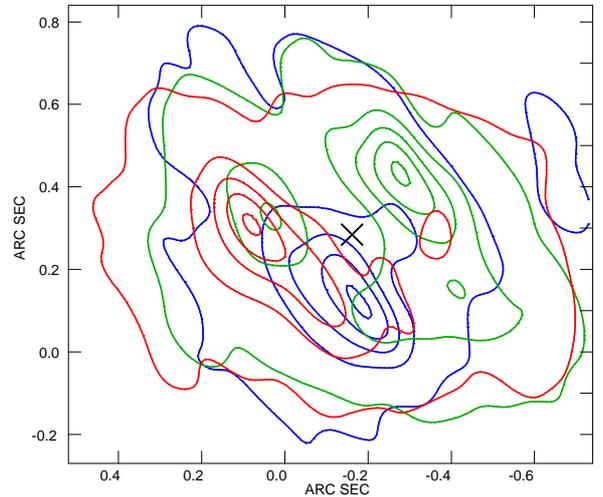}
\vspace*{-0.2cm}
\caption{325 GHz masers, summed over $<$17, 17--27, and $>$27 km
  s$^{-1}$, shown in blue, green, and red. Contour levels start at 30
  Jy; higher levels are at 24\%, 48\%, 72\%, and 96\% of the peak,
  where the peak is 2184, 1018, and 903 Jy for the red, green, and blue
  ranges. These levels have been chosen to emphasise the brighter
  emission and indicate the total extent at a uniform sensitivity
  limit (to 1--3\% peak), omitting fine details. The cross marks
  {\bf VY.} }
         \label{VY325_ALL.CASSQA.PS}
   \end{figure}
}

   \begin{figure}
   
   \includegraphics[width=10cm]{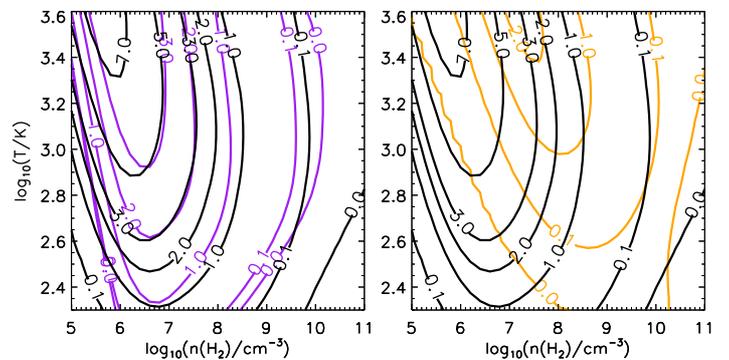}
\vspace*{-0.5cm}
\caption{Maser (negative) optical depths for 22 GHz (black, both
  panels), 325 GHz masers (magenta, left panel) and 321 GHz (orange,
  right panel). Amplification occurs if this exceeds zero.}
         \label{Neufeldmasers_vycma.eps}
   \end{figure}


\section{Discussion}

{\bf VY} coincides with the maser centre of expansion and is almost
certainly the location of the star. This was suspected
(\citealt{Muller07}; \citealt{Kaminski13}), but never before resolved
as a distinct continuum peak.  Our
position (Sect.~\ref{Continuum}) agrees within the uncertainties
with the centre of expansion of SiO and 22 GHz
H$_2$O masers (\citealt{Zhang12}; \citealt{Choi08}).

Exponential maser amplification exaggerates small differences in
conditions, so single-epoch data are interpreted with caution.  The
658 GHz masers have a complex distribution close to the star, but
mostly outside the typical SiO outer rim, taken from
\citet{Richter13}, as shown in Fig.~\ref{VYCMa_321325658_ravel.eps}
(although \citet{Shinnaga04} found a more extended bipolar
outflow). The asymmetric emission at large angular separations and
expansion velocities (Fig.~\ref{VYCMa_321325658_ravel.eps}) is
inconsistent with a simple velocity-radius relationship.  The inner
658 GHz masers arise in the pulsation-dominated region, but straddle
the dust formation zone identified by \citet{Decin06} and indicated by
the 22 GHz inner radius.

The patchy ring of bright 321 and 325 GHz masers within $\sim$10 km
s$^{-1}$ of $V_{\star}$ (Fig.~\ref{321-325xyplots}) is typical of
tangential beaming from a radially accelerating shell. 
 The 321 GHz
masers are undetectable towards the star, suggesting that they trace
particularly strongly accelerated gas.
The red- and blue-shifted emission
at large angular separations suggests additional, complex motions.
The brightest 325 GHz masers occur at velocities close to
$V_{\star}$, seen in Fig.~\ref{VY325_ALL.CASSQA.PS}, but asymmetries
are seen in the south and east offsets of the moderately blue- and
red-shifted peaks.
Figures~\ref{321-325xyplots} and~\ref{658xyplots} suggest that the
masers, especially at 658 GHz, trace shocks where the stellar wind
encounters continuum peak {\bf C}.  
Elongated, bright maser features characteristic of shocks, seen at up
to $\sim$150 mas from the star, could be associated with dust formation
\citep{Bladh13} or wind collisions arising from inhomogenous mass
loss/speeds \citep{Zijlstra01}.   

We grouped the
maser components into features  and attempted to
cross-match the clumps of different transitions, but found no
significant associations, so co-propagation of these sub-mm
masers seems unlikely.   A comparison with 22 GHz was not made because of the
difference in epochs; around similar evolved stars these are concentrated in clumps
with a filling factor $\la$1\%, containing 30--90\% of the total
mass loss in this region \citep{Richards12}, embedded in less dense gas.


The lines studied arise from distinct, although overlapping zones, and
are segregated on scales of a few tens of au.  All are thought to
  be collisionally pumped \citep{Yates97}.
  Figure~\ref{Neufeldmasers_vycma.eps} compares the physical conditions
  required for masing of the three vibrational ground-state lines 
using an
  excitation model similar to that described by \citet{Neufeld91} 
upated as detailed in Appendix~\ref{sec:mmodel}.  The
conditions in the VY CMa CSE lie around the locus from top right (near
the star) to bottom left, but with a variation of up to $\times70$ in
number density and  $\times0.7$ in temperature in denser
clumps or compressed material.  The contours mark the negative optical
depths in the direction of the velocity gradient for unsaturated
masers; the optical depths in other directions may greatly exceed the
values shown here (optical depths of 22 GHz masers have been estimated at
up to $\sim$13 in RSG S Per, \citealt{Richards11}). The 321 GHz gain
drops rapidly below 1000 K, whilst the 22 and 325 GHz masers extend
to cooler temperatures. The 325 GHz maser is collisionally quenched at
lower densities than the 22 GHz and 321 GHz transitions.

  The models of
\citet{Humphreys01} predicted a comparable radial range for 22
 and
325 GHz masers around low-mass stars, the latter being favoured in
cooler, more rarefied conditions, possibly in a two-phase
medium. These predictions agree well with the observed
inner radii at succesively higher distances from the star at 321, 22,
and 325 GHz, if the inner radii of 321 and 325 GHz masers are
determined by conditions in the inter-clump gas surrounding the 22 GHz
clouds.
Not yet published models (Gray) require temperatures $>$1500 K and H$_2$
number densities $\sim$ 6$\times$10$^9$ cm$^{-3}$ for 658 GHz masers,
likely to be found in the inner parts of its observed distribution,
but the more extended emission is surprising.  The excitation of
321 GHz masers at 500 mas (600 au) from the star is also puzzling
since typical temperatures of 250 K \citep{Decin06} are too low, so
local heating might be important.


\section{Summary}
These ALMA observations have identified the centre of expansion of
high-excitation masers with a distinct continuum peak, VY CMa itself,
360 au NW of the brightest sub-mm dust emission.  The 658, 321, and
325 GHz masers are found at increasing distances from the star as
predicted, but reach unexpectedly large separations. The
high-excitation 658 and possibly 321 GHz masers cross the dust
formation zone and some emission appears to emanate from shocked
regions surrounding the star and tracing the limbs of {\bf C};
interaction with the uniquely complex dust distribution is undoubtedly
significant. The different transitions form clumps that do not
overlap even when found at similar separations from {\bf VY}. The
velocities are generally consistent with expansion but deviate
drastically and irregularly from spherical symmetry.  We will compare
the maser kinematics with models of a flared disc plus bipolar
outflow (\citealt{Decin06}; \citealt{Muller07}) and ejection along
less symmetric arcs as described by \citet{Humphreys07}.  
These
observations provide the first opportunity to test sub-mm maser models
rigorously,
to be followed-up in future papers along with an analysis of the
kinematics, thermal lines, and dust.


\begin{acknowledgements} {This paper makes use of ALMA data:
    ADS/JAO.ALMA $\#$2011.0.00011.SV.  ALMA is a partnership of ESO
    (representing its member states), NSF (USA) and NINS (Japan),
    together with NRC (Canada) and NSC and ASIAA (Taiwan), in
    cooperation with the Republic of Chile. The Joint ALMA Observatory
    is operated by ESO, AUI/NRAO and NAOJ. We thank the anonymous
    referee for thoughtful comments that have greatly improved the
    clarity of this paper.}
\end{acknowledgements}


\Online
\begin{appendix} 
\label{appendix}

\section{Water maser lines}
Table~\ref{tab:masers} gives some properties of the sub-mm maser lines resolved
by ALMA and the well-known 22 GHz maser.
\begin{table}
\caption{H$_2$O masers. $\nu_2$: vibrational state; spin: ortho/para isomers. }
\begin{tabular}{rclrcc}
\hline  \hline 
\multicolumn{1}{c}{Frequency} &Transition &$\nu_2$  &$E_U$ & Spin & Discovery\\
\multicolumn{1}{c}{(GHz)} &($J_{Ka,Kc}$)&&(K) & 
(level) &(reference)\\
\hline
22.23508&$6_{1,6}$--$5_{2,3}$&0&643&$o$&C69\\
321.22564&$10_{2,9}$--$9_{3,6}$&0&1862&$o$&M90\\
325.15292& $5_{1,5}$--$4_{2,2}$&0&470&$p$&M91\\
658.00655& $1_{1,0}$--$1_{0,1}$&1&2361&$o$&M95\\
\hline
\end{tabular}
\tablebib{C69 Cheung et al. (1969); M90 Menten et al. (1990); M91 Menten \& Melnick (1991); M95 Menten \& Young (1995).}
\label{tab:masers}
\end{table}

\section{Observations and data processing}
VY CMa was observed by ALMA on 2013 16--19 August using 16--20 12 m
antennas.  The primary objective of these Science Verification (SV)
observations was to demonstrate the ability to observe on baselines up
to 2.7 km and develop calibration techniques involving strong, narrow
spectral lines.  Three separate configurations or scheduling blocks (SBs),
covering each of the maser lines, are referred to as the 321
GHz,
325 GHz, and 658 GHz SBs; details are given in Table~\ref{tab:obs}.
These were divided into one or more spectral windows (spw)
covering $\sim$850--1700 km s$^{-1}$ (depending on SB). 
Each spw was divided into 3840 channels, but as a result of Hanning smoothing
in the correlator, the finest effective velocity resolution is
approximately double the channel spacing, that is, 0.45 km s$^{-1}$ for
321 and 658 GHz, and 0.9 km s$^{-1}$ at 325 GHz.  The data and scripts
(including a description of the procedures) used for calibration and
initial imaging are available from {\small
  http://almascience.eso.org/alma-data/science-verification}. As a
Science Verification project, some observational methods were
experimental, for example the duration of phase-referencing cycles
turned out to allow a few phase ambiguities at the highest
frequencies. Methods such as band-to-band phase transfer and fast
switching will be available in future. These observations required
very dry atmospheric conditions, so the weather determined their
duration within the time available for Science Verification.  Normal
ALMA calibration and imaging procedures were followed, using {\sc
  casa} ({\small http://casa.nrao.edu}).  Each SB was executed
  three times at different hour angles, giving a total time on VY CMa at
each frequency of $\sim$1.5 hr in addition to calibration
observations. The phase-reference source J0648-3044 was observed in
1.5 min scans, bracketing 6.75 min on VY CMa for the 321 GHz SB and
5.25 min for the 325 and 658 GHz SBs.

\begin{table}
\caption{Observational parameters. $\Delta_{\nu\,\mathrm{cont}}$:
  effective continuum bandwidth for the whole SB, giving noise $\sigma_{\mathrm{rms\,cont}}$.  $\delta{\nu}$: unaveraged
  channel spacing; $\sigma_{\mathrm{rms1}}$:
  noise in 1 km s$^{-1}$ channels free from bright emission.}
\small
\begin{tabular}{lcccccc}
\hline \hline
SB  $\!\!\!$&spw centre$\!\!\!$& spw width $\!\!\!\!\!\!$&$\Delta_{\nu\,\mathrm{cont}}$$\!\!\!\!\!\!$&$\sigma_{\mathrm{rms\,cont}}$  $\!\!\!\!\!\!$&$\delta{\nu}$ $\!\!\!$&$\sigma_{\mathrm {rms1}}$\\
(GHz) $\!\!\!$& (GHz)    $\!\!\!$& (GHz) $\!\!\!\!\!\!$& (GHz) $\!\!\!\!\!\!$& (mJy)   $\!\!\!\!\!\!$&(kHz) $\!\!\!$& (mJy) \\
\hline
321 $\!\!\!$& 321.18305$\!\!\!$& 0.9375$\!\!\!\!\!\!$&1.74 $\!\!\!\!\!\!$&0.31 $\!\!\!\!\!\!$&244.141     $\!\!\!$&2.5\\
321 $\!\!\!$& 322.44385$\!\!\!$& 0.9375$\!\!\!\!\!\!$& $\!\!\!\!\!\!$&       $\!\!\!\!\!\!$&244.141    $\!\!\!$&2.5\\
321 $\!\!\!$& 310.95845$\!\!\!$& 0.9375$\!\!\!\!\!\!$& $\!\!\!\!\!\!$&       $\!\!\!\!\!\!$&244.141    $\!\!\!$&1.6\\
321 $\!\!\!$& 309.95865$\!\!\!$& 0.9375$\!\!\!\!\!\!$& $\!\!\!$&       $\!\!\!\!\!\!$&244.141  $\!\!\!$&1.7\\
325 $\!\!\!$& 325.10740$\!\!\!$& 1.873$\!\!\!\!\!\!$&2.25 $\!\!\!\!\!\!$&1.1  $\!\!\!\!\!\!$&488.281     $\!\!\!$& 7.5\\
325 $\!\!\!$& 321.96140$\!\!\!$& 1.873$\!\!\!\!\!\!$& $\!\!\!\!\!\!$&       $\!\!\!\!\!\!$&488.281     $\!\!\!$& 1.8\\
658 $\!\!\!$& 657.82750$\!\!\!$& 1.872$\!\!\!\!\!\!$&0.4$\!\!\!\!\!\!$&2.4   $\!\!\!\!\!\!$&488.281   $\!\!\!$& 20\\
\hline
\end{tabular}
\label{tab:obs}
\end{table}
Antenna positions were updated where required, and applied corrections
derived from system temperature and water vapour radiometry
measurements. The precipitable water vapour (PWV) was 0.3 mm except
for the last of the three 658 GHz observations, when it was 0.7 mm. The
water vapour radiometry corrections produced very significant
improvements, especially for the 658 SB taken at 0.7 mm PWV. A small
amount of bad data were excised.

The bright QSO J0522-3627 was used for bandpass calibration.  This was
observed for the default duration of 5.25 min in each of the three
executions of the 321 GHz SB.  However, it was only observed for 2.5
min in each of the 325 GHz and 658GHz SBs.  After all calibration was
complete, we checked the variation of the imaged continuum emission with
frequency within each SB.  The channel-to-channel position scatter was
as expected from the signal-to-noise ratio (S/N), without any systematic position shift, and
the flux density was consistent with the expected spectral index
$\sim$2, so we are satisfied that the bandpass does not lead to
misleading results. The main symptom was that the noise rms decreased
more shallowly than the expected inverse square-root dependence on
the number of channels averaged.  The position uncertainties were
  also affected by dynamic range limitations in imaging and by
possibly incomplete modelling of the atmosphere in the deep water-absorption lines.

Pallas was used as the primary flux
scale calibrator  (Butler-JPL-Horizons 2012, ALMA Memo 594), selecting 
baselines shorter than the first null in the visibilities. Using the
321 GHz SB,  the flux density derived from Pallas for the
phase-reference source, J0648-3044, was $0.433\pm0.008$ Jy at
reference  frequency 316.093 GHz, spectral index $\alpha$ $-0.80\pm0.03$.
Since the 325 GHz data covered similar frequencies but had
a poorer S/N  than the 321-GHz data, we extrapolated the 321
GHz values
to the relevant frequencies for the 325 GHz data. 
At 658 GHz, the flux density of J0648-3044 derived from
Pallas is 0.28 Jy, compared with $0.24\pm0.02$ Jy
extrapolated from 321 GHz.  This may not be a fair comparison, since
there is no guarantee that the spectral index is linear from 321 to
658 GHz, but it suggests that the error could be up to 15\%.

The phase-reference source, J0648-3044, 9$^{\circ}$ from VY CMa, was
used to derive time-dependent phase and amplitude corrections. The
phase could be  connected smoothly between successive scans for most
antennas and times, but in a few cases where there was an ambiguity,
the target scan affected was excluded from the initial imaging.

After applying instrumental and calibration source corrections, the VY
CMa data in each SB was adjusted to constant velocity with respect to the Local Standard
of Rest (LSR). All velocities are given as $V_{\mathrm{LSR}}$.
Low-resolution cubes were made for each data set to identify
line-free continuum, and we made preliminary images to check
the astrometry.  In each SB, the brightest maser channel was identified
and imaged, providing a starting model for self-calibration.  After
several iterations, the
solutions were applied to all channels.  The
solutions were applied to all channels and to the data initially
excluded because of the phase-referencing ambiguities noted above.

The bandwidth corresponding to the sum of line-free channels (spread
over the whole observing bandwidth) $\Delta_{\nu\,\mathrm{cont}}$ and the
image noise rms $\sigma_{\mathrm{rms\,cont.}}$ are given in
Table~\ref{tab:obs}. The mean frequencies were 316 and 319 GHz for the
data sets referred to as 321 and 325 GHz.
The continuum
channels were imaged using natural weighting, which gave a synthesised beam of
(0\farcs22$\times$0\farcs13) at 321 and 325 GHz, and
(0\farcs11$\times$0\farcs06) at 658 GHz. In all cases the beam
position angle (PA) was $\sim$28$^{\circ}$.

The shortest baseline was 14 m, and inspection of the visibility
amplitudes against baseline length shows that the flux density remains quite
steady out to 70 m at 658 GHz and 170 m at 321--325 GHz, suggesting
that we recover all the flux on scales $<6$'' or $<13$'' at the higher
or lower frequencies.  We compared the total continuum flux densities
with literature values (Fu et al 2012; Kami\'{n}ski et al. 2013;
Muller et al. 2007; Shinnaga et a. 2004).  All the measurements using
$\sim1''$ aperture lie close to a spectral index of $2.2\pm0.2$. Those taken
using a larger aperture, such as at 658-GHz from Shinnaga et al. (2004) and
by Knapp \& Woodhams (1993) using the JCMT (effective aparture $\sim18$''), are
higher, for instance $0.62\pm0.04$ Jy at 240 GHz, $2.18\pm0.24$ Jy at 353 GHz
and $9.7\pm1.5$ Jy at 677 GHz.  This suggests that there is an
extended component of dust on scales larger than we sampled.


The  continuum was subtracted from each data set and partial uniform
weighting (Briggs weighting with robust=0.5 as defined by {\sc casa}) was used to image the masers,
giving beam sizes of (0\farcs18$\times$0\farcs09) and
(0\farcs088$\times$0\farcs044) at 321/325 and 658 GHz, respectively.
No spectral averaging was applied, so the measurements for each
maser channel are not completely independent as a result of the Hanning
smoothing in the correlator. All image extents were
$\le$80\% of the primary beam, so no primary beam correction was
applied.

The maser peak brightnesses and S/N in each of the cubes were
321 GHz: 426.6 Jy beam$^{-1}$, S/N 2010; 325 GHz: 271 Jy beam$^{-1}$,
S/N 1330; 658 GHz: 361 Jy beam$^{-1}$, and S/N 764.  In the maser line
wings (not dynamic-range limited but affected by the atmosphere) the
$\sigma_{\mathrm{rms}}$ noise values were 4, 15, and 40 mJy for 321,
325, and 658 GHz, respectively. The values for all spw at lower
resolution are given in Table~\ref{tab:obs}.  

We measured the
positions of the masers and continuum peaks by fitting two-dimensional
Gaussian components using the {\sc aips} task {\sc sad}. We did
  not attempt to resolve the individual components since the smallest
  beam size $\sim50$ mas is much larger than the probable maser beamed
  size, although this might be possible for the brightest masers. Thus
all flux densities are measured over the restoring beam.
 The relative
position uncertainties are given by (beam size)/S/N (for fairly sparse
$uv$ coverage in narrow channels (Condon et al. 1998; Richards 1997).  We selected components
$>$$3\sigma_{\mathrm{rms}}$ at 321 and 325 GHz or
$>$$4\sigma_{\mathrm{rms}}$ at 658 GHz (where $\sigma_{\mathrm{rms}}$
was measured separately off-source for each channel) and rejected those that
obviously coincided with sidelobes. We rejected components that did not
form series of at least three in successive channels, within the
maximum position uncertainty.

The total flux in fitted maser components at 321, 325, and 658 GHz is
95\%, 65\%, and 76\% of the integrated map flux density for each SB.
However, the fraction of flux recovered in components is no higher for
channels containing peaks $>$100 Jy, implying that the main loss is
due to deconvolution errors putting power into sidelobes, since the
325  and 658 GHz masers are more affected by atmospheric
conditions. The component selection method  avoids
locating spurious positions, at the expense of loss of peak flux.
We grouped the
maser components into features comprising series of components in
successive channels within the position errors, as in Richards et al. (2012), and attempted to
cross-match the clumps of different transitions but found no significant associations. In each case, 5--10\% of features have
pairs within 50 mas, 2 km s$^{-1}$ , but applying a 50 mas shift to one
data set produces a similar number of pairs, so this seems like random
coincidence.

\section{Maser modelling}
\label{sec:mmodel}

Figure~\ref{Neufeldmasers_vycma.eps} compares the physical conditions
needed to excite the observed maser transitions (at 22 GHz and 321 and
325 GHz) within the ground-vibrational state.  Here, we present
results obtained with an excitation model that included the combined
effects of collisional excitation by H$_2$, spontaneous radiative
decay, and radiative trapping of infrared transitions. We adopted the
latest quantal rate coefficients (Daniel et al. 2011) for
collisionally induced transitions amongst the lowest 45 rotational
states of ortho- and para-H$_2$O, together with an extrapolation
(Neufeld 2010) to the next 75 rotational states. We treated the
effects of radiative trapping with the use of an escape probability
method and the assumption of a steep velocity gradient in a single
direction (e.g. Neufeld \& Melnick 1991).  The results plotted here are for
an effective water column density, $N({\mathrm {H_2O}}) = n({\mathrm {H_2O}})
/ (dv/dz)$ of 10$^{17}$~cm$^{-2}$~per~km~s$^{-1}$; based upon the density and
velocity profiles obtained for VY CMa by Decin et al. (2006), this
value is appropriate for the general outflow at distances in the range
150 to 750 au, corresponding to angular offsets of 125 to 625 mas (the
region within which most of the maser spots are located).  The
plotted contours show, as a function of temperature
and H$_2$ density, the negative optical depths predicted in the
direction of the velocity gradient.  These optical depths were computed in the
unsaturated limit, where the population inversion is assumed to be
undiminished by the effects of stimulated emission.

The contours labelled zero mark the boundary of the region within which
the level populations are inverted.  That region covers a broad range
of densities and temperatures for all three transitions considered,
although the 321 GHz maser gain drops rapidly below $\sim 1000$~K, as
expected given its relatively high upper state energy.  Similar
calculations, not presented here, for the 658~GHz transition show a
similar behaviour; this transition, too, shows a significant maser gain
only at high temperature.  Clearly, in the limit of high density, the
population inversion for any transition inevitably disappears as the
level populations approach LTE.  However, the quenching density
above which the population inversion vanishes varies from transition
to transition and is clearly lowest for the 325 GHz transition.  This
behaviour may explain why the 325 GHz spots, as plotted in Figure 5
(purple squares), have a larger inner boundary than the other masing
transitions.

Despite the large overlap of the regions in parameter space within
which strong maser amplification can occur
(Fig.~\ref{Neufeldmasers_vycma.eps}), there are few or no exact
coincidences between the 321 and 325 GHz maser spots observed
simultaneously.  As noted previously, this may simply reflect the tendency
of the exponential amplification process to accentuate small
differences in opacity.


\begin{thebibliography}{33}
\expandafter\ifx\csname natexlab\endcsname\relax\def\natexlab#1{#1}\fi

\bibitem[{{Alcolea} {et~al.}(2013){Alcolea}, {Bujarrabal}, {Planesas},
  {Teyssier}, {Cernicharo}, {De Beck}, {Decin}, {Dominik}, {Justtanont}, {de
  Koter}, {Marston}, {Melnick}, {Menten}, {Neufeld}, {Olofsson}, {Schmidt},
  {Sch{\"o}ier}, {Szczerba}, \& {Waters}}]{Alcolea13}
{Alcolea}, J., {Bujarrabal}, V., {Planesas}, P., {et~al.} 2013, \aap, 559, A93

\bibitem[{{Bladh} {et~al.}(2013){Bladh}, {H{\"o}fner}, {Nowotny}, {Aringer}, \&
  {Eriksson}}]{Bladh13}
{Bladh}, S., {H{\"o}fner}, S., {Nowotny}, W., {Aringer}, B., \& {Eriksson}, K.
  2013, \aap, 553, A20

\bibitem[{Bowen(1988)}]{Bowen88}
Bowen, G.~H. 1988, ApJ, 329, 299

\bibitem[{{Chiavassa} {et~al.}(2011){Chiavassa}, {Freytag}, {Masseron}, \&
  {Plez}}]{Chiavassa11}
{Chiavassa}, A., {Freytag}, B., {Masseron}, T., \& {Plez}, B. 2011, \aap, 535,
  A22

\bibitem[{{Choi} {et~al.}(2008){Choi}, {Hirota}, {Honma}, {Kobayashi},
  {Bushimata}, {Imai}, {Iwadate}, {Jike}, {Kameno}, {Kameya}, {Kamohara},
  {Kan-Ya}, {Kawaguchi}, {Kijima}, {Kim}, {Kuji}, {Kurayama}, {Manabe},
  {Maruyama}, {Matsui}, {Matsumoto}, {Miyaji}, {Nagayama}, {Nakagawa},
  {Nakamura}, {Oh}, {Omodaka}, {Oyama}, {Sakai}, {Sasao}, {Sato}, {Sato},
  {Shibata}, {Tamura}, {Tsushima}, \& {Yamashita}}]{Choi08}
{Choi}, Y.~K., {Hirota}, T., {Honma}, M., {et~al.} 2008, \pasj, 60, 1007

\bibitem[{{Daniel} \& {Cernicharo}(2013)}]{Daniel13}
{Daniel}, F. \& {Cernicharo}, J. 2013, \aap, 553, A70

\bibitem[{{Decin} {et~al.}(2006){Decin}, {Hony}, {de Koter}, {Justtanont},
  {Tielens}, \& {Waters}}]{Decin06}
{Decin}, L., {Hony}, S., {de Koter}, A., {et~al.} 2006, \aap, 456, 549

\bibitem[{{Gray}(2012)}]{Gray12}
{Gray}, M. 2012, {Maser Sources in Astrophysics} (Cambridge University Press)

\bibitem[{Habing(1996)}]{Habing96}
Habing, H.~J. 1996, A\&A Rev., 7, 97

\bibitem[{{Hartmann} \& {MacGregor}(1980)}]{Hartmann80}
{Hartmann}, L. \& {MacGregor}, K.~B. 1980, \apj, 242, 260

\bibitem[{{Humphreys} {et~al.}(2001){Humphreys}, {Yates}, {Gray}, {Field}, \&
  {Bowen}}]{Humphreys01}
{Humphreys}, E.~M.~L., {Yates}, J.~A., {Gray}, M.~D., {Field}, D., \& {Bowen},
  G.~H. 2001, A\&A, 379, 501

\bibitem[{{Humphreys} {et~al.}(2007){Humphreys}, {Helton}, \&
  {Jones}}]{Humphreys07}
{Humphreys}, R.~M., {Helton}, L.~A., \& {Jones}, T.~J. 2007, \aj, 133, 2716

\bibitem[{{Hunter} {et~al.}(2007){Hunter}, {Young}, {Christensen}, \&
  {Gurwell}}]{Hunter07}
{Hunter}, T.~R., {Young}, K.~H., {Christensen}, R.~D., \& {Gurwell}, M.~A.
  2007, in IAU Symposium, ed. J.~M. {Chapman} \& W.~A. {Baan}, Vol. 242,
  481--488

\bibitem[{{Ireland} \& {Scholz}(2006)}]{Ireland06}
{Ireland}, M.~J. \& {Scholz}, M. 2006, \mnras, 367, 1585

\bibitem[{{Kami{\'n}ski} {et~al.}(2013){Kami{\'n}ski}, {Gottlieb}, {Young},
  {Menten}, \& {Patel}}]{Kaminski13}
{Kami{\'n}ski}, T., {Gottlieb}, C.~A., {Young}, K.~H., {Menten}, K.~M., \&
  {Patel}, N.~A. 2013, \apjs, 209, 38

\bibitem[{{Lipscy} {et~al.}(2005){Lipscy}, {Jura}, \& {Reid}}]{Lipscy05}
{Lipscy}, S.~J., {Jura}, M., \& {Reid}, M.~J. 2005, \apj, 626, 439

\bibitem[{{Mauron} \& {Josselin}(2011)}]{Mauron11}
{Mauron}, N. \& {Josselin}, E. 2011, \aap, 526, A156

\bibitem[{{Muller} {et~al.}(2007){Muller}, {Dinh-V-Trung}, {Lim}, {Hirano},
  {Muthu}, \& {Kwok}}]{Muller07}
{Muller}, S., {Dinh-V-Trung}, {Lim}, J., {et~al.} 2007, \apj, 656, 1109

\bibitem[{Neufeld \& Melnick(1991)}]{Neufeld91}
Neufeld, D.~A. \& Melnick, G.~J. 1991, ApJ, 368, 215

\bibitem[{{Norris} {et~al.}(2012){Norris}, {Tuthill}, {Ireland}, {Lacour},
  {Zijlstra}, {Lykou}, {Evans}, {Stewart}, \& {Bedding}}]{Norris12}
{Norris}, B.~R.~M., {Tuthill}, P.~G., {Ireland}, M.~J., {et~al.} 2012, Nat.,
  484, 220

\bibitem[{{O'Gorman} {et~al.}(2014){O'Gorman}, {Vlemmings}, {Richards},
  {Baudry}, {De Beck}, {Decin}, {Harper}, {Humphreys}, {Kervella}, {Khouri}, \&
  {Muller}}]{OGorman14}
{O'Gorman}, E., {Vlemmings}, W., {Richards}, A.~M.~S., {et~al.} 2014, ArXiv
  e-prints

\bibitem[{{Patel} {et~al.}(2007){Patel}, {Curiel}, {Zhang}, {Sridharan}, {Ho},
  \& {Torrelles}}]{Patel07}
{Patel}, N.~A., {Curiel}, S., {Zhang}, Q., {et~al.} 2007, \apjl, 658, L55

\bibitem[{{Richards} {et~al.}(2011){Richards}, {Elitzur}, \&
  {Yates}}]{Richards11}
{Richards}, A.~M.~S., {Elitzur}, M., \& {Yates}, J.~A. 2011, A\&A, 525, A56

\bibitem[{{Richards} {et~al.}(2012){Richards}, {Etoka}, {Gray}, {Lekht},
  {Mendoza-Torres}, {Murakawa}, {Rudnitskij}, \& {Yates}}]{Richards12}
{Richards}, A.~M.~S., {Etoka}, S., {Gray}, M.~D., {et~al.} 2012, A\&A, 546, A16

\bibitem[{Richards {et~al.}(1998)Richards, Yates, \& Cohen}]{Richards98V}
Richards, A. M.~S., Yates, J.~A., \& Cohen, R.~J. 1998, MNRAS, 299, 319

\bibitem[{{Richter} {et~al.}(2013){Richter}, {Kemball}, \& {Jonas}}]{Richter13}
{Richter}, L., {Kemball}, A., \& {Jonas}, J. 2013, \mnras, 436, 1708

\bibitem[{{Shinnaga} {et~al.}(2004){Shinnaga}, {Moran}, {Young}, \&
  {Ho}}]{Shinnaga04}
{Shinnaga}, H., {Moran}, J.~M., {Young}, K.~H., \& {Ho}, P.~T.~P. 2004, \apjl,
  616, L47

\bibitem[{{Wittkowski} {et~al.}(2007){Wittkowski}, {Boboltz}, {Ohnaka},
  {Driebe}, \& {Scholz}}]{Wittkowski07}
{Wittkowski}, M., {Boboltz}, D.~A., {Ohnaka}, K., {Driebe}, T., \& {Scholz}, M.
  2007, A\&A, 470, 191

\bibitem[{{Wittkowski} {et~al.}(2012){Wittkowski}, {Hauschildt},
  {Arroyo-Torres}, \& {Marcaide}}]{Wittkowski12}
{Wittkowski}, M., {Hauschildt}, P.~H., {Arroyo-Torres}, B., \& {Marcaide},
  J.~M. 2012, \aap, 540, L12

\bibitem[{{Woitke}(2006)}]{Woitke06}
{Woitke}, P. 2006, A\&A, 460, L9

\bibitem[{Yates {et~al.}(1997)Yates, Field, \& Gray}]{Yates97}
Yates, J.~A., Field, D., \& Gray, M.~D. 1997, MNRAS, 285, 383

\bibitem[{{Zhang} {et~al.}(2012){Zhang}, {Reid}, {Menten}, \&
  {Zheng}}]{Zhang12}
{Zhang}, B., {Reid}, M.~J., {Menten}, K.~M., \& {Zheng}, X.~W. 2012, \apj, 744,
  23

\bibitem[{Zijlstra {et~al.}(2001)Zijlstra, te~Lintel~Hekkert, Chapman, Likkel,
  Comeron, Norris, Molster, \& Cohen}]{Zijlstra01}
Zijlstra, A.~A., te~Lintel~Hekkert, P., Chapman, J.~M., {et~al.} 2001, MNRAS,
  322, 280

\end{thebibliography}

\vspace*{5cm}
\makeatletter \renewcommand\@seccntformat[1]{} \makeatother 
\section{References (Appendix)}
{
\tiny
\expandafter\ifx\csname natexlab\endcsname\relax\def\natexlab#1{#1}\fi
Cheung, A.~C., Rank, D.~M., Townes, C.~H., Thornton, D.~D., \& Welch, W.~J.
\hspace*{0.35cm}1969, Nat, 221, 626\\
Condon, J.~J., Cotton, W.~D., Greisen, E.~W., {et~al.} 1998, AJ, 115, 1693\\
{Daniel}, F., {Dubernet}, M.-L., \& {Grosjean}, A. 2011, \aap, 536, A76\\
{Decin}, L., {Hony}, S., {de Koter}, A., {et~al.} 2006, \aap, 456, 549\\
{Fu}, R.~R., {Moullet}, A., {Patel}, N.~A., {et~al.} 2012, \apj, 746, 42\\
{Kami{\'n}ski}, T., {Gottlieb}, C.~A., {Young}, K.~H., {Menten}, K.~M., \&
  {Patel}, N.~A. 2013, \hspace*{0.35cm}\apjs, 209, 38\\
Knapp, G.~R. \& Woodhams, M. 1993, in Massive Stars: \hspace*{0.1cm}Their  \hspace*{0.1cm}Lives  \hspace*{0.1cm}in  \hspace*{0.1cm}the  \hspace*{0.35cm}Interstellar Medium, ed. J.~P. Cassinelli \& E.~B. Churchwell, Vol.~35 (ASP
  \hspace*{0.35cm}conference series, San Francisco), 199\\
Menten, K.~M. \& Melnick, G.~J. 1991, ApJ, 377, 647\\
Menten, K.~M., Melnick, G.~J., \& Phillips, T.~G. 1991, ApJ, 350, L41\\
Menten, K.~M. \& Young, K. 1995, ApJ, 450, L67\\
{Muller}, S., {Dinh-V-Trung}, {Lim}, J., {et~al.} 2007, \apj, 656, 1109\\
{Neufeld}, D.~A. 2010, \apj, 708, 635\\
Neufeld, D.~A. \& Melnick, G.~J. 1991, ApJ, 368, 215\\
Richards, A. M.~S. 1997, PhD thesis, University of Manchester\\
{Richards}, A.~M.~S., {Etoka}, S., {Gray}, M.~D., {et~al.} 2012, A\&A, 546, A16\\
{Shinnaga}, H., {Moran}, J.~M., {Young}, K.~H., \& {Ho}, P.~T.~P. 2004, \apjl,
  616, L47
}


\end{appendix}
\end{document}